\documentclass{article}
\usepackage{amssymb}
\usepackage{amsfonts}
\usepackage{amsmath}
\usepackage{latexsym}

\begin{document}

\title{Artificial Wormhole}
\author{ A. A. Kirillov E.P. Savelova \\
\emph{Dubna International University of Nature, Society and Man,}\\
\emph{Universitetskaya Str. 19, Dubna, 141980, Russia } }
\date{}
\maketitle

\begin{abstract}
It is shown that recently reported result by the OPERA Collaboration \cite%
{light} of an early arrival time of muon neutrinos with respect to the speed
of light in vacuum does not violate standard physical laws. We show that
vacuum polarization effects in intensive external fields may form a
wormhole-like object. The simplest theory of such an effect is presented and
basic principles of formation of an artificial wormhole are also considered.
\end{abstract}

1. An actual wormhole requires the violation of averaged null energy
condition (ANEC) Ref. \cite{Vis} and is commonly supposed to be forbidden in
classical physics. However, the situation changes when we consider a virtual
wormhole. It represents a quantum topology fluctuation which takes place at
very small (Plankian) scales and lasts for a very short period of time \cite%
{wheeler,H78}. It does not obey the Einstein equations and, therefore, ANEC
cannot forbid the origin of such an object. By other words, virtual
wormholes violate readily ANEC. Moreover, virtual wormholes play an
important role in particle physics. First, they may introduce in a natural
way the cutoff at very small scales and remove divergencies in quantum field
theory \cite{KS10}. Secondly, they predict new phenomena and open new
perspectives in applied physics. In particular, by applying an external
field one may govern the intensity of such fluctuations and thus organize an
artificial wormhole. It is curious that actual traversable wormholes may be
supported by arbitrarily small quantities of \textquotedblleft exotic
matter\textquotedblright\ Ref. \cite{vis2}, while in our case the latter is
represented by virtual wormholes.

We note that such an artificial wormhole is an extremely complex object
(from microphysics standpoint) which requires numerical study. In the
present paper we consider the simplest model which introduces an anisotropy
in the speed of light which allows to give the principle explanation of the
anomaly observed by the OPERA Collaboration \cite{light} (faster than light
travel of muon neutrinos). We also point out to Ref. \cite{sar} where
another approach has been suggested to explain the OPERA anomaly.

2. In what follows we widely use results of \cite{KS10} and will not repeat
them here. A virtual wormhole can be described as follows. It is convenient
from the very beginning to use the Euclidean approach (e.g., see Refs. \cite%
{H78}, \cite{bab}-\cite{hwu} and references therein). Then the simplest
virtual wormhole is described by the metric ($\alpha =1,2,3,4$)%
\begin{equation}
ds^{2}=h^{2}\left( r\right) \delta _{\alpha \beta }dx^{\alpha }dx^{\beta },
\label{wmetr}
\end{equation}%
where
\begin{equation}
h\left( r\right) =1+\theta \left( a-r\right) \left( \frac{a^{2}}{r^{2}}%
-1\right)
\end{equation}%
and $\theta \left( x\right) $ is the step function. Such a wormhole has
vanishing throat length. Indeed, in the region $r>a$, $h=1$ and the metric
is flat, while the region $r<a$, with the obvious transformation $y^{\alpha
}=\frac{a^{2}}{r^{2}}x^{\alpha }$, is also flat for $y>b$. Therefore, the
regions $r>a$ and $r<a$ represent two Euclidean spaces glued at the surface
of a sphere $S^{3}$ with the center at the origin $r=0$ and radius $r=b$.
Such a space can be described with the ordinary double-valued flat metric in
the region $r_{\pm }>a$ by
\begin{equation}
ds^{2}=\delta _{\alpha \beta }dx_{\pm }^{\alpha }dx_{\pm }^{\beta },
\label{wmetr2}
\end{equation}%
where the coordinates $x_{\pm }^{\alpha }$ describe two different sheets of
space. Now, identifying the inner and outer regions of the sphere $S^{3}$
allows the construction of a wormhole which connects regions in the same
space (instead of two independent spaces). This is achieved by gluing the
two spaces in (\ref{wmetr2}) by motions of the Euclidean space (the Poincare
motions). If $R_{\pm }$ is the position of the sphere in coordinates $x_{\pm
}^{\mu }$, then the gluing is the rule%
\begin{equation}
x_{+}^{\mu }=R_{+}^{\mu }+\Lambda _{\nu }^{\mu }\left( x_{-}^{\nu
}-R_{-}^{\nu }\right) ,  \label{gl}
\end{equation}%
where $\Lambda _{\nu }^{\mu }\in O(4)$, which represents the composition of
a translation and a rotation of the Euclidean space (Lorentz
transformation). In terms of common coordinates such a wormhole represents
the standard flat space in which the two spheres $S_{\pm }^{3}$ (with
centers at positions $R_{\pm }$) are glued by the rule (\ref{gl}). We point
out that the physical region is the outer region of the two spheres. Thus,
in general, the wormhole is described by a set of parameters: the throat
radius $b$, positions of throats $R_{\pm }$ , and rotation matrix $\Lambda
_{\nu }^{\mu }\in O(4)$.

3. Consider now the simplest massless scalar field and construct the Green
function in the presence of a gas of virtual wormholes.

The Green function obeys the Laplace equation
\begin{equation*}
-\Delta G\left( x,x^{\prime }\right) =4\pi ^{2}\delta \left( x-x^{\prime
}\right)
\end{equation*}%
with proper boundary conditions at throats (we require $G$ and $\partial
G/\partial n$ to be continual at throats). The Green function for the
Euclidean space is merely $G_{0}\left( x,x^{\prime }\right) =\frac{1}{\left(
x-x^{\prime }\right) ^{2}}$ (and $G_{0}\left( k\right) =4\pi ^{2}/k^{2}$ for
the Fourier transform). In the presence of a single wormhole which connects
two Euclidean spaces this equation admits the exact solution. For outer
region of the throat $S^{3}$ the source $\delta \left( x-x^{\prime }\right) $
generates a set of multipoles placed in the center of sphere which gives the
corrections to the Green function $G_{0}$ in the form (we suppose the center
of the sphere at the origin)%
\begin{equation}
\delta G=-\frac{1}{x^{2}}\sum_{n=1}^{\infty }\frac{1}{n+1}\left( \frac{a}{%
x^{\prime }}\right) ^{2n}\left( \frac{x^{\prime }}{x}\right) ^{n-1}Q_{n},
\end{equation}%
where $Q_{n}=\frac{4\pi ^{2}}{2n}\sum_{l=0}^{n-1}\sum_{m=-l}^{l}Q_{nlm}^{%
\ast \prime }Q_{nlm}$ and $Q_{nlm}\left( \Omega \right) $ are
four-dimensional spherical harmonics e.g., see Ref. \cite{fock}. In the
present paper we shall consider a dilute gas approximation and, therefore,
it is sufficient to retain the lowest (monopole) term only. A single
wormhole which connects two regions in the same space is a couple of
conjugated spheres $S_{\pm }^{3}$ of the radius $a$ with a distance $\vec{X}=%
\vec{R}_{+}-\vec{R}_{-}$ between centers of spheres. So the parameters of
the wormhole are\footnote{%
The additional parameter (rotation matrix $U$) is important only for
multipoles of higher orders.} $\xi =(a,R_{+},R_{-})$. The interior of the
spheres is removed and surfaces are glued together. Then the proper boundary
conditions (the actual topology) can be accounted for by adding the bias of
the source
\begin{equation}
\delta (x-x^{\prime })\rightarrow N\left( x,x^{\prime }\right) =\delta
(x-x^{\prime })~+b\left( x,x^{\prime }\right) .
\end{equation}%
In the approximation $a/X\ll 1$ (e.g., see for details \cite{KS07}) the bias
takes the form
\begin{equation}
b_{0}\left( x,x^{\prime },\xi \right) =\frac{a^{2}}{2}\left( \frac{1}{\left(
R_{-}-x^{\prime }\right) ^{2}}-\frac{1}{\left( R_{+}-x^{\prime }\right) ^{2}}%
\right) \left[ \delta ^{4}(x-R_{+})-\delta ^{4}(x-R_{-})\right] .  \label{b1}
\end{equation}%
We expect that virtual wormholes have throats $a\sim \ell _{pl}$ of the
Plankian size, while in the present paper we are interested in much larger
scales. Therefore, the form (\ref{b1}) is sufficient for our aims. However
this form is not acceptable in considering the short-wave behavior and
vacuum polarization effects (stress energy tensor). In the last case one
should account for the finite value of the throat size and replace in (\ref%
{b1}) the point-like source with the surface density (induced on the throat)
i.e., see for details \cite{KS10}, $\delta ^{4}(x-R_{\pm })\rightarrow \frac{%
1}{2\pi ^{2}a^{3}}\delta (\left\vert x-R_{\pm }\right\vert -a).$

In the rarefied gas approximation the bias function for the gas of wormholes
is additive, i.e.,
\begin{equation}
b_{total}\left( x,x^{\prime }\right) =\sum b_{0}\left( x,x^{\prime },\xi
_{i}\right) =N\int b_{0}(x,x^{\prime },\xi )F(\xi )d\xi ,
\end{equation}%
Where
\begin{equation}
F\left( \xi \right) =\frac{1}{N}\sum\limits_{i=1}^{N}\delta \left( \xi -\xi
_{i}\right) .  \label{F}
\end{equation}

We assume a homogeneous but anisotropic distribution $F(\xi )=F\left(
a,X\right) $, then for the bias we find
\begin{equation}
b_{total}\left( x-x^{\prime }\right) =\int a^{2}\left( \frac{1}{R_{-}^{2}}-%
\frac{1}{R_{+}^{2}}\right) \delta ^{4}(x-x^{\prime }-R_{+})NF\left(
a,X\right) d\xi  \label{bx}
\end{equation}%
Consider the Fourier transform $F\left( a,X\right) =\int F\left( a,k\right)
e^{-ikX}\frac{d^{4}k}{\left( 2\pi \right) ^{4}}$ and using the integral $%
\frac{1}{x^{2}}=\int \frac{4\pi ^{2}}{k^{2}}e^{-ikx}\frac{d^{4}k}{\left(
2\pi \right) ^{4}}$ we find for $b\left( k\right) =\int b\left( x\right)
e^{ikx}d^{4}x$ the expression
\begin{equation}
b_{total}\left( k\right) =N\int a^{2}\frac{4\pi ^{2}}{k^{2}}\left( F\left(
a,k\right) -F\left( a,0\right) \right) da.  \label{bk}
\end{equation}%
Consider now a particular form for $F\left( a,X\right) $, e.g.,
\begin{equation}
NF\left( a,X\right) =n\delta \left( a-a_{0}\right) \frac{1}{2}\left( \delta
^{4}\left( X-r_{0}\right) +\delta ^{4}\left( X+r_{0}\right) \right) ,
\label{NF}
\end{equation}%
where $n=N/V$ is the density of wormholes. Such a distribution corresponds
to a coherent set of wormholes with the throat $a_{0}$, oriented along the
same direction $r_{0}$ and with the distance between throats $%
r_{0}=\left\vert R_{+}-R_{-}\right\vert $. We assume that $r_{0}=(0,\vec{r}%
_{0})$ has only spatial direction. Then $NF\left( a,k\right) =\int NF\left(
a,X\right) e^{ikx}d^{4}x$ reduces to $NF\left( a,k\right) =n\delta \left(
a-a_{0}\right) \cos \left( \vec{k}~\vec{r}_{0}\right) $. Thus from (\ref{bk}%
) we find
\begin{equation}
b\left( k\right) =-na^{2}\frac{4\pi ^{2}}{k^{2}}\left( 1-\cos \left( \vec{k}%
\text{ }\vec{r}_{0}\right) \right) .  \label{B}
\end{equation}

In the vacuum case the background fluctuations have an isotropic and
homogeneous distribution and form the background cutoff function $\overline{N%
}\left( k\right) $, so that the regularized vacuum Green function $%
G_{reg}\left( k\right) $ has the form%
\begin{equation}
G_{reg}\left( k\right) =\overline{N}\left( k\right) G_{0}\left( k\right) =%
\frac{4\pi ^{2}}{k^{2}}\overline{N}\left( k\right) .  \label{GF}
\end{equation}%
General properties of the cutoff is that $\overline{N}\left( k\right)
\rightarrow 0$ as $k\gg k_{pl}$ and $\overline{N}\left( k\right) \rightarrow
const\ll 1$ on the mass shell (as \thinspace $k\ll k_{pl}$). We shall use
Plankian units. i.e., $k_{pl}=2\pi $.

4. Consider the structure of the bias of the unit source (\ref{B}) in the
coordinate representation. Substituting (\ref{NF}) into (\ref{bx}) we find%
\begin{equation}
b\left( x\right) =-\frac{na^{2}}{2}\left( \frac{2}{x^{2}}-\frac{1}{\left(
x+r_{0}\right) ^{2}}-\frac{1}{\left( x-r_{0}\right) ^{2}}\right) .
\end{equation}%
We recall that here $\frac{1}{\left( x-x^{\prime }\right) ^{2}}=G_{0}\left(
x,x^{\prime }\right) $ is the standard Euclidean Green function which, upon
the continuation to the Minkowsky space, transforms to the Feynman
propagator which is important in quantum field theory. However when
considering the propagation of signals we should use the retarding Green
function $G_{0}$ $\rightarrow $ $G_{ret}\left( x,x^{\prime }\right) =\frac{1%
}{R}\delta (t^{\prime }-t+\frac{1}{c}R)$, while the bias has the same
structure (e.g., see Ref. \cite{KSS09}). Thus we see that the additional
source represents three outgoing spherical waves which originate at
positions $x=0$ and $x=\pm r_{0}$. Since $r_{0}$ has only spatial direction
the additional source $b\left( x\right) $ forms the wavefront which overruns
the standard wave in the direction $\vec{r}_{0}$ which should lead to the
observed anomaly $\Delta t=r_{0}/c$. The intensity of such an additional
signal is described by the portion of the primary signal scattered on
virtual wormholes which is given by $b=-\int b\left( x\right) d^{4}x$%
\begin{equation}
b=2\pi ^{2}na^{2}r_{0}^{2}\ll 1.
\end{equation}%
We recall that the dilute gas approximation requires $2\pi ^{2}na^{4}\ll 1$
(i.e., the portion of the volume cut by virtual wormholes should be
sufficiently small), while the ratio $r_{0}/a$ may be an arbitrary parameter
and therefore in general $b$ may reach the order of unity.

5. Consider now the generating functional (the partition function) which is
used to generate all possible correlation functions in quantum field theory
(and the perturbation scheme when we include interactions)
\begin{equation}
Z_{total}\left( J\right) =\sum\limits_{\tau }\sum\limits_{\varphi }e^{-S_{E}}
\end{equation}%
where the sum is taken over field configurations $\varphi $ and topologies $%
\tau $ (wormholes), the Euclidean action is
\begin{equation}
S_{E}=-\frac{1}{2}\left( \varphi \Delta \varphi \right) +4\pi ^{2}\left(
J\varphi \right) ,  \label{act}
\end{equation}%
and we use the notions $\left( J\varphi \right) =\int J\left( x\right)
\varphi \left( x\right) d^{4}x$. Here $J$ denotes an external current. The
sum over field configurations $\varphi $ can be replaced by the integral
\begin{equation}
Z^{\ast }\left( J\right) =\int \left[ D\varphi \right] e^{\frac{1}{2}\left(
\varphi \Delta \varphi \right) -\left( J\varphi \right) }.  \label{gf1}
\end{equation}%
Upon the simple transformations
\begin{equation}
\frac{1}{2}\left( \varphi \Delta \varphi \right) -\left( J\varphi \right) =%
\frac{1}{2}\left( \widetilde{\varphi }\Delta \widetilde{\varphi }\right) -%
\frac{1}{2}\left( JGJ\right) ,
\end{equation}%
where $\widetilde{\varphi }=\varphi -GJ$ and $G~$is the background Green
function (\ref{GF}), we cast the partition function to the form
\begin{equation}
Z^{\ast }=\int \left[ D\widetilde{\varphi }\right] e^{\frac{1}{2}\left(
\widetilde{\varphi }\Delta \widetilde{\varphi }\right) -\frac{1}{2}\left(
JGJ\right) }=Z_{0}(G)e^{-\frac{1}{2}\left( JGJ\right) },  \label{gf2}
\end{equation}%
where $Z_{0}(G)=\int \left[ D\varphi \right] e^{\frac{1}{2}\left( \varphi
\Delta \varphi \right) }$ is the standard expression and $G=G\left( \xi
_{1},...,\xi _{N}\right) $ is the Green function for a fixed topology, i.e.,
for a fixed set of wormholes $\xi _{1},...,\xi _{N}$ .

Consider now the sum over topologies $\tau $. To this end we restrict with
the sum over the number of wormholes and integrals over parameters of
wormholes:
\begin{equation}
\sum\limits_{\tau }\rightarrow \sum\limits_{N}\int
\prod\limits_{i=1}^{N}d\xi _{i}=\int \left[ DF\right]  \label{ts}
\end{equation}%
where $F$ is given by (\ref{F}). We point out that in general the
integration over parameters is not free (e.g., it obeys the obvious
restriction $\left\vert \vec{R}_{i}^{+}-\vec{R}_{i}^{-}\right\vert \geq
2a_{i}$). This defines the generating function as
\begin{equation}
Z_{total}\left( J\right) =\int \left[ DF\right] Z_{0}(G)e^{-\frac{1}{2}%
\left( JGJ\right) }.
\end{equation}%
Since in the vacuum case virtual wormholes have a homogeneous distribution,
in the Fourier representation the bias $N(x,x^{\prime },\xi )\rightarrow
N(k,k^{\prime },\xi )$ which gives $N(k,k^{\prime })$ $=N(k,\xi )\delta
(k-k^{\prime })$, then we find compare with (\ref{GF})
\begin{equation*}
G\left( k\right) =G_{0}\left( k\right) N(k,\xi ).
\end{equation*}%
Then for the total partition function we find
\begin{equation}
Z_{total}\left( J\right) =\int \left[ DN(k)\right] e^{-I(N(k))}e^{-\frac{1}{2%
}\sum \frac{4\pi ^{2}}{k^{2}}\left\vert J_{k}\right\vert ^{2}N(k)},
\label{ztot}
\end{equation}%
where $\sum_{k}=\frac{L^{4}}{\left( 2\pi \right) ^{4}}\int d^{4}k$ and $%
\left[ DN\right] =\prod\limits_{k}dN_{k}$ . The functional $I(N)$ comes from
the integration measure (which includes the Jacobian of transformation from $%
F\left( \xi \right) $ to $N\left( k\right) $)
\begin{equation*}
e^{-I(N)}=\int \left[ DF\right] Z_{0}(N(k,\xi ))\delta \left( N\left(
k\right) -N\left( k,\xi \right) \right)
\end{equation*}%
and has the sense of the action for the bias function $N\left( k\right) $.
In the true vacuum case $J=0$ and by means of using the expression (\ref%
{ztot}) we find the two-point Green function in the form
\begin{equation}
G\left( k\right) =\frac{4\pi ^{2}}{k^{2}}\overline{N}(k)  \label{grfm}
\end{equation}%
where $\overline{N}(k)$ is the cutoff function (the mean bias) which is
given by
\begin{equation*}
\overline{N}(k)=\frac{1}{Z_{total}\left( 0\right) }\int \left[ DN\right]
e^{-I\left( N\right) }N\left( k\right) .
\end{equation*}

The action $I(N)$ can be expanded as\footnote{%
In this expansion the cutoff $\overline{N}(k)$ is merely the solution of $%
\frac{\delta I\left( N\right) }{\delta N\left( k\right) }=0$.}
\begin{equation}
I(N)=I(\overline{N})+\frac{1}{2}\sum_{k}\frac{\left( N\left( k\right) -%
\overline{N}(k)\right) ^{2}}{\sigma _{k}^{2}}+...  \label{actn}
\end{equation}%
where $\sigma _{k}^{2}$ defines the dispersion of vacuum topology
fluctuations\footnote{%
We point out that the inverse dispersion $\sigma _{k}^{-2}$ is analogous to
the Laplace operator $\Delta $ in (\ref{act}).}. Since the bias $N\left(
k\right) $ plays the role of a projection operator \cite{KS10} which for a
dense gas (e.g., see (\ref{B})) has the asymptotic $\overline{N}\left(
k\right) \rightarrow const\ll 1$as \thinspace $k\ll k_{pl}$ one may expect $%
\sigma _{k}^{2}=\overline{N}(k)(1-\overline{N}(k))\simeq \overline{N}(k)$ as
\thinspace $k\ll k_{pl}$.

6. Consider now topology fluctuations in the presence of an external
current. In the presence of an external current $J^{ext}$ the intensity of
topology fluctuations changes. Indeed using (\ref{ztot}), (\ref{actn}) we
find
\begin{equation}
I(N,J^{ext})=I(\overline{N})+\frac{1}{2}\sum_{k}\frac{\left( N\left(
k\right) -\overline{N}(k)\right) ^{2}}{\sigma _{k}^{2}}+\frac{1}{2}\sum_{k}%
\frac{4\pi ^{2}}{k^{2}}\left\vert J_{k}^{ext}\right\vert ^{2}N(k)+...
\end{equation}%
which gives
\begin{equation}
I(N,J^{ext})=I(\overline{N},J^{ext})+\frac{1}{2}\sum_{k}\frac{\left( N\left(
k\right) -(\overline{N}(k,J^{ext})\right) ^{2}}{\sigma _{k}^{2}}
\end{equation}%
where
\begin{equation}
I(\overline{N},J^{ext})=I(\overline{N})-\frac{1}{2}\sum_{k}\sigma
_{k}^{2}\left( \frac{4\pi ^{2}}{2k^{2}}\left\vert J_{k}^{ext}\right\vert
^{2}\right) ^{2}
\end{equation}%
and
\begin{equation}
\overline{N}(k,J^{ext})-\overline{N}(k)=b\left( J\right) =-\sigma _{k}^{2}%
\frac{4\pi ^{2}}{2k^{2}}\left\vert J_{k}^{ext}\right\vert ^{2}
\end{equation}%
Since we expect that the external current has scales $k\ll k_{pl}$, (where $%
\frac{\sigma _{k}^{2}}{\overline{N}(k)}\simeq 1$) this expression can be
cast into the form
\begin{equation}
b\left( J\right) =-\frac{1}{2}\frac{\sigma _{k}^{2}}{\overline{N}(k)}\frac{%
4\pi ^{2}}{k^{2}}\overline{N}(k)\left\vert J_{k}^{ext}\right\vert ^{2}\simeq
-\frac{1}{2}G_{reg}\left( k\right) \left\vert J_{k}^{ext}\right\vert ^{2}
\end{equation}%
where $G=G_{reg}$ is the physically measured (observed at laboratory scales)
Green function. At very large scales upon renormalization of charge values
we get $G_{reg}=G_{0}=$ $\frac{4\pi ^{2}}{k^{2}}$. Now comparing this
function with (\ref{B}) we relate the additional distribution of virtual
wormholes and the current as%
\begin{equation}
N\int a^{2}\frac{4\pi ^{2}}{k^{2}}\left( F\left( a,0\right) -F\left(
a,k\right) \right) da\simeq \frac{1}{2}G_{reg}\left( k\right) \left\vert
J_{k}^{ext}\right\vert ^{2}.
\end{equation}%
which gives at scales $k\ll k_{pl}$%
\begin{equation}
\left\vert J_{k}^{ext}\right\vert ^{2}\simeq 2n\overline{a^{2}}\left(
f\left( 0\right) -f\left( k\right) \right) .
\end{equation}%
where $n\overline{a^{2}}f\left( k\right) =$ $N\int a^{2}F\left( a,k\right)
da $, or for a particular distribution (\ref{B}) the necessary current to
produce the desired additional fluctuations takes the form%
\begin{equation}
\left\vert J_{k}^{ext}\right\vert ^{2}\simeq 2na_{0}^{2}\left( 1-\cos \left(
\vec{k}\text{ }\vec{r}_{0}\right) \right) =\frac{b}{\pi ^{2}r_{0}^{2}}\left(
1-\cos \left( \vec{k}\text{ }\vec{r}_{0}\right) \right)  \label{J}
\end{equation}%
Here $a_{0}$ is the typical size of the throat of virtual wormholes and $b$
is the portion of the signal scattered on topology ($b<1$).

7. Thus, we see that sufficiently intensive external current (which probably
was reached in the experiment \cite{light} ) is apt to produce the anomaly
observed (faster than light travel). In conclusion we also note that we
considered here the simplest situation which does not destroy the
homogeneity of space. However it is clear that our consideration allows for
the straightforward generalization on a more complex (e.g., spherically
symmetric) case which opens the new perspective to create such an object in
laboratory physics. Moreover, external intensive fields are widely met in
astrophysics, and therefore, one may expect that such complex objects
(actual wormhole-like objects) are indeed responsible for the dark matter
phenomenon, e.g., see \cite{KS11} which however requires for the further
studying.

\end{document}